\documentclass[aps,pra,twocolumn,showpacs, superscriptaddress]{revtex4-1}

\usepackage{graphicx}
\usepackage{color}
\usepackage[usenames,dvipsnames]{xcolor}

\newcommand{\nc}{\newcommand}
\nc{\eq}{\begin{equation}}
\nc{\eeq}{\end{equation}}
\nc{\eqa}{\begin{eqnarray}}
\nc{\eeqa}{\end{eqnarray}}
\nc{\ar}{\begin{array}}
\nc{\ear}{\end{array}}
\nc{\bfig}{\begin{figure}}
\nc{\efig}{\end{figure}}
\nc{\dg}{\dagger}
\nc{\eps}{\frac{\epsilon}{2}}
\nc{\juuri}{\sqrt{\Omega^2+(\eps)^2}}
\nc{\sx}{\sigma_x}
\nc{\sy}{\sigma_y}
\nc{\sz}{\sigma_z}
\nc{\spl}{\sigma_+}
\nc{\sm}{\sigma_-}
\nc{\Sx}{\bar{\sigma}_x}
\nc{\Sy}{\bar{\sigma}_y}
\nc{\Sz}{\bar{\sigma}_z}
\nc{\Spl}{\bar{\sigma}_+}
\nc{\Sm}{\bar{\sigma}_-}
\nc{\nn}{\nonumber}
\nc{\noi}{\noindent}
\nc{\omt}{\tilde{\omega}}
\nc{\Somt}{S(\omt)}
\nc{\Somtd}{S^{\dg}(\omt)}
\nc{\got}{\gamma_{\omega}(t)}
\nc{\gmot}{\gamma_{-\omega}(t)}
\nc{\po}{\mathcal{P}}
\nc{\qo}{\mathcal{Q}}
\nc{\adg}{a^{\dg}}
\nc{\gammat}{\tilde{\gamma}}
\nc{\Q}{$\mathcal{Q }$}
\nc{\C}{$\mathcal{C }$}
\nc{\kvec}{\mathbf{k}}

\begin{document}

\title{Coherence trapping and information back-flow in dephasing qubits}

\author{Carole Addis}
\affiliation{SUPA, EPS/Physics, Heriot-Watt University, Edinburgh, EH14 4AS, UK}
\email{ca99@hw.ac.uk}
\author{Gregoire Brebner}
\affiliation{SUPA, EPS/Physics, Heriot-Watt University, Edinburgh, EH14 4AS, UK}
\author{Pinja Haikka}
\email[]{pinjahaikka@gmail.com}
\affiliation{Department of Physics and Astronomy, Aarhus University, DK-8000 Aarhus C, Denmark.}
\author{Sabrina Maniscalco} 
\affiliation{SUPA, EPS/Physics, Heriot-Watt University, Edinburgh, EH14 4AS, UK}
\affiliation{Turku Center for Quantum Physics, Department of Physics and Astronomy, University of Turku, FIN-20014 Turku, Finland}
\email[]{s.maniscalco@hw.ac.uk} 

\date{\today}

\begin{abstract}
We study the interplay between coherence trapping, information back-flow and the form of the reservoir spectral density for dephasing qubits. We show that stationary coherence is maximized when the qubit undergoes non-Markovian dynamics, and we elucidate the different roles played by the low and high frequency parts of the environmental spectrum. We show that the low frequencies fully determine the presence or absence of information back-flow while the high frequencies dictate the maximal amount of coherence trapping.
\end{abstract}


\maketitle

The ability to manipulate coherently qubit systems both individually and collectively is one of the key pre-requisites of quantum technologies. This has been achieved in a number of physical systems including laser cooled trapped ions, atoms in optical lattices, nitrogen-vacancy centers in diamonds and quantum dots \cite{lasercooled, opticallattices, NVdiamond, dots}. Generally, due to the interaction with the environment, the qubits will be subjected to decoherence and dissipation phenomena whose specific timescales and characteristics strongly depend on the physical context considered. In all of the systems above, there is a clear distinction between the decoherence (dephasing) time scale $T_2$ and the dissipation (heating) time scale $T_1$, the latter one being considerably longer than the former one. This means that the major limiting source of environmental noise in such systems can be described, for times $t \lesssim T_1$, as pure dephasing.

Long-lasting electronic coherences in biological surroundings and the formation of steady state entanglement in coherently coupled dimer systems have been shown to be crucially linked to the presence of non-Markovian noise \cite{Nature, Susana}. Moreover, non-Markovian environments have been found to be an important resource in fundamental quantum processes such as quantum metrology \cite{Susana2}, quantum key distribution \cite{Ruggero}, teleportation \cite{Teleportation} and quantum communication \cite{Bogna}. 

In this paper we investigate dephasing dynamics, and explore the interplay between the ability of a single qubit to partly retain coherences in the long time limit, that is, the phenomenon of coherence trapping, and the presence of information back-flow due to reservoir memory effects. 

In the spirit of reservoir engineering we consider the following questions: What are the conditions that optimize stationary coherence of a dephasing qubit? Are memory effects able to improve the robustness of the qubits to environmental noise? How does the form of the reservoir spectral density affect such robustness? Our aim is to better understand the physical mechanisms leading to coherence trapping in a paradigmatic open quantum system model \cite{luczka, massimo, reina}, by linking this phenomenon to properties such as reservoir memory and the form of the spectral density function. As the latter quantity is experimentally modifiable (see, e.g., Refs. \cite{luczka}-\cite{us}), this study holds significance in the design of noise-robust quantum-enhanced devices.

{\it The model.}
Let us consider an exactly solvable model of pure dephasing (setting $\hbar = 1$):
 \begin{eqnarray}
H= \omega_0 \sigma_z+ \sum_k  \omega_k a^{\dag}_k a_k + \sum_k  \sigma_z (g_k a_k+ g_k^* a^{\dag}_k), \nonumber
 \end{eqnarray}
where $\omega_0$ is the qubit frequency, $\omega_k$ the frequencies of the reservoir modes, $a_k \;(a_k^{\dag})$ the annihilation (creation) operators and $g_k$ the coupling constant between each reservoir mode and the qubit. In the continuum limit $\sum_k |g_k|^2 \rightarrow \int d\omega J(\omega) \delta (\omega_k-\omega)$, where $J(\omega)$ is the reservoir spectral density. It is worth noticing that the interaction Hamiltonian commutes with the qubits Hamiltonian but not with the field Hamiltonian. Hence, due to the finite interaction energy, the state of the field evolves even if initially it was at zero temperature. This in turn causes the (pure) decoherence of the qubit.

The evolution of the coherences of a single qubit is given by  $\rho_{ij} (t) = e^{-\Lambda(t)} \rho_{ij} (0),\, i\neq j$, while the diagonal elements remain invariant under the effect of the environmental noise \cite{luczka, massimo, reina}. The dephasing factor is
\begin{eqnarray} \label{factor}
 \Lambda(t) &=& 2 \int_0^{\infty} d \omega\, g(\omega,T)[1-\cos(\omega t)],\label{eq:Gamma} \nonumber\\
 g(\omega,T)&=&\frac{J(\omega)}{\omega^2}\coth\left(\frac{\omega}{2k_B T}\right), 
\end{eqnarray}
where $T$ is the temperature of the environment which is assumed to be in a thermal state. Note that the $ g(\omega,T)$ functions, in the limiting cases of zero and high temperatures, are connected through the relation $g(\omega, \text{high-T}) = 2 k_B T g(\omega,0) / \omega$. 

The details of the qubit dynamics are fully dictated by the spectral density function characterising the system-environment interaction. Depending of the physical realization of the purely dephasing model the spectral density function can take several different forms, and its specific structure can, in some cases, be modified by reservoir engineering techniques \cite{probes}. This implies a way to control the dynamical features of the qubit in a way that does not involve direct interaction with the qubit, like one would do if using, for example, dynamical decoupling or bang bang-control techniques \cite{bang1,bang2}. In the following we study how different properties of the spectral density function are related to dynamical features such as coherence trapping and back-flow of information.

{\it The spectral density.}
A common class of spectral density functions extensively used in the literature is the family of Ohmic spectra \cite{J},
\begin{eqnarray}
J(\omega)=  \frac{\omega^s}{\omega_c^{s-1}} f(\omega, \omega_c ), \label{eq:Jomega}
\end{eqnarray}
parametrized by a real positive number $s$. By changing the $s$-parameter in Eq. (\ref{eq:Jomega}), it is possible to go from sub-Ohmic reservoirs ($s<1$) to Ohmic ($s=1$) and super-Ohmic ($s>1$) reservoirs, respectively, by controlling the strength of interaction of the low-frequency part of the spectrum. We stress that such engineering of the Ohmicity of the spectrum is possible, e.g., when simulating the dephasing model in trapped ultracold atoms, as demonstrated in Ref. \cite{us}. The high-frequency part of the spectrum, instead, is controlled by the cutoff function $f(\omega, \omega_c )$, with $\omega_c$ the cutoff frequency \cite{J}. By definition, the cutoff function does not affect the low-frequency part of the spectrum, i.e., $J(\omega) \simeq \omega^s / \omega_c ^{s-1}$, for $\omega/\omega_c \ll 1$, but it suppresses the high frequency contribution such that $J(\omega) \simeq 0$ for $\omega/\omega_c \gg 1$. 

For the sake of concreteness we compare two types of functions, both of exponential form but featuring a softer or harder cutoff with respect to one another. Such softer and harder cutoff functions are typical, e.g., of dephasing quantum dots \cite{Lorent}, and they take the form
\eq
f_{\text{soft}}(\omega,\omega_c)=e^{-\omega/\omega_c},\;
f_{\text{hard}}(\omega,\omega_c)=e^{-(\omega/\omega_c)^2},
\eeq
respectively. A comparison between the two will elucidate the role of the high frequency modes in the qubit dynamics; for a given cutoff frequency $\omega_c$, the harder cutoff function strongly suppresses the high frequency modes in comparison to the softer cutoff function. 

{\it Coherence trapping.}
Markovian models of dephasing are characterized by exponential decay of the qubit coherences, i.e., $\rho_{ij} (t) = e^{-\lambda t } \rho_{ij} (0)$, where $\lambda$ is a constant decay factor, hence predicting vanishing coherences in the long time limit. The situation is different, however, for the exact dephasing model here considered. Depending on the specific form of the spectral density, the decoherence factor $\Lambda(t)$ can either diverge asymptotically or reach a positive non-zero value. In  the former case no coherences survive, while in the latter case qubit dephasing will stop after a finite time, therefore leading to coherence trapping. 

The asymptotic divergence of the decoherence factor $\Lambda(t)$ depends only on the value of $\omega g(\omega,T)$ in the limit $\omega \rightarrow 0$ \cite{tomipinja}. More precisely, for $t \rightarrow \infty$, $\Lambda(t)$ diverges when $  \omega  g(\omega,T)$ diverges in the origin. On the contrary $\Lambda(t)$ has a finite asymptotic value when $\omega g(\omega,T) $ vanishes in the origin \cite{note}.  This behavior is independent on the specific form of the cutoff function, provided that it is finite at $\omega=0$ and it is sufficiently well-behaving. To appreciate this point, it is useful to look at the properties of the dephasing rate $\gamma(t)$, defined as the time derivative of $\Lambda (t)$: 
\begin{eqnarray} \label{pinja}
&& \gamma(t)=\int d\omega J(\omega)\coth[\omega/2k_BT]\sin(\omega t)/\omega. \label{eq:Gamma}
\end{eqnarray}
With the Ohmic class of spectral density functions it is straightforward to check that, for any temperature $T$, $\omega g(\omega,T)$ diverges in the origin for $s \le 1$ and that the dephasing rate has a finite limit $\lim_{t\rightarrow\infty}\gamma(t)=\lambda>0$. In this limit the dephasing is effectively Markovian and all coherences are lost.

On the other hand, for $s>1$, the integrand of the dephasing rate is sufficiently regular to allow to approximate the long time behavior as $\gamma (t \rightarrow \infty) \approx J(0) =0 $, where we have used the fact that $\lim_{t\rightarrow \infty} \sin (\omega t) / \omega = \pi \delta(\omega)$. This means that the dephasing rate, for these values of $s$, converges to zero, stopping dephasing and hence causing coherence trapping. While the specific form of the cutoff function does not affect the presence or absence of coherence trapping, as we will see in the following, the weight of the high frequency part of the spectrum does influence the value of the stationary coherences in the case when they are present. 

\begin{figure}[h]
\includegraphics[width=0.5\textwidth]{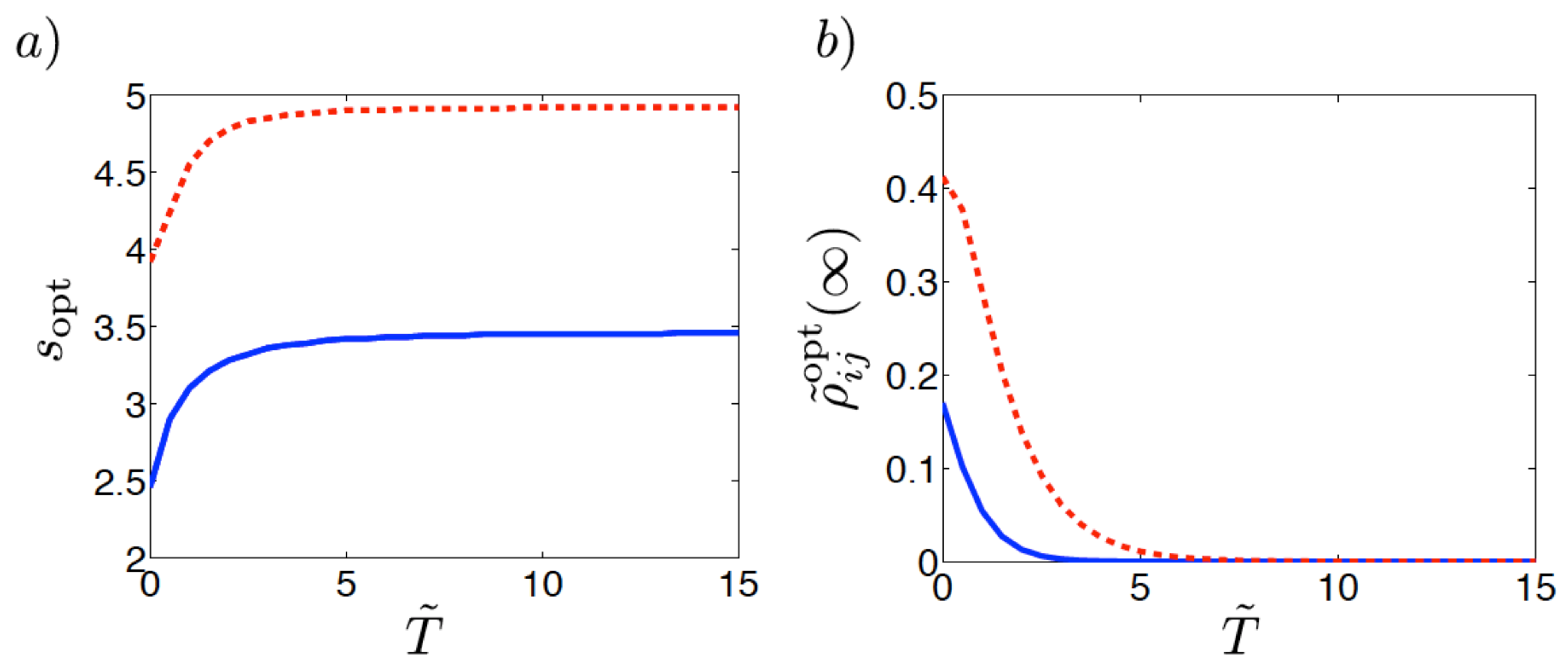}
\caption{(Color online) Temperature dependence $\tilde{T}$ of a) $s_{\text{opt}}$ and b) the stationary coherences $\tilde{\rho}_{ij}(\infty)$ for the softer (blue solid line) and harder (red dotted line) cutoff functions.  Here, $\tilde{T}=2k_BT/\omega_c$ is a dimensionless temperature. } 
\label{figure1}
\end{figure} 

We will now explore the value of Ohmicity parameter leading to the maximum stationary coherences for $s > 1(2)$ in the zero (high) $T$ temperature regimes. Throughout, we define stationary coherence in terms of the initial state, i.e., $\tilde{\rho}_{ij}(\infty)=\rho_{ij} (\infty)/\rho_{ij} (0)$. For the softer cutoff function the stationary coherences take a simple analytical expression in the zero $T$ and high $T$ limits, $\tilde{\rho}_{ij} (\infty)= e^{- 2 \Gamma (s-1)}$ ($s \ge 1$), and  $\tilde{\rho}_{ij} (\infty)  = e^{- 2  \tilde{T}  \Gamma (s-2)}$ ($s \ge 2$), respectively, with $\Gamma(x)$ the Euler Gamma function. Here we have defined a dimensionless temperature, $\tilde{T}=2k_BT/\omega_c$. From these expressions it is easy to derive the values of Ohmicity parameter $s_{\text{opt}}$ maximizing the stationary coherences: $s_{\text{opt}} \simeq 2.46$ for zero $T$ and $s_{\text{opt}} \simeq 3.46$ for high $T$. On the other hand, for the harder cutoff function, in the zero $T$ and high $T$ limits, $\tilde{\rho}_{ij}(\infty) = e^{- \Gamma (\frac{1}{2}(-1+s))}$ ($s \ge 1$) and $\tilde{\rho}_{ij}(\infty) = e^{- \tilde{T}\Gamma (-1+\frac{s}{2})}$ ($s \ge 2$).  In this case, $s_{\text{opt}} \simeq 3.92$ for zero $T$ and $s_{\text{opt}} \simeq 4.92$ for high $T$. In general both $s_{\text{opt}}$ and the maximum stationary coherence will depend on temperature as shown in Fig. 1 (a) and (b). As temperature increases, there is a sharp increase in $s_{\text{opt}}$ as it converges to a stationary value for high temperatures. In the long time limit, the coherences are increasingly destroyed as temperature is increased.

In Fig. 2 we compare the stationary coherences as a function of $s$, in the zero-$T$ case, for the softer and harder cutoff. The figure clearly shows that the harder cutoff function leads to a more efficient coherence trapping as the values of the stationary coherences are higher than those obtained for the softer cutoff, for any value of $s$. This can be understood noting that the qubit interacts with less reservoir modes in the harder cutoff case, mitigating the overall effect of dephasing.
\begin{figure}[h]   
\includegraphics[width=0.3\textwidth]{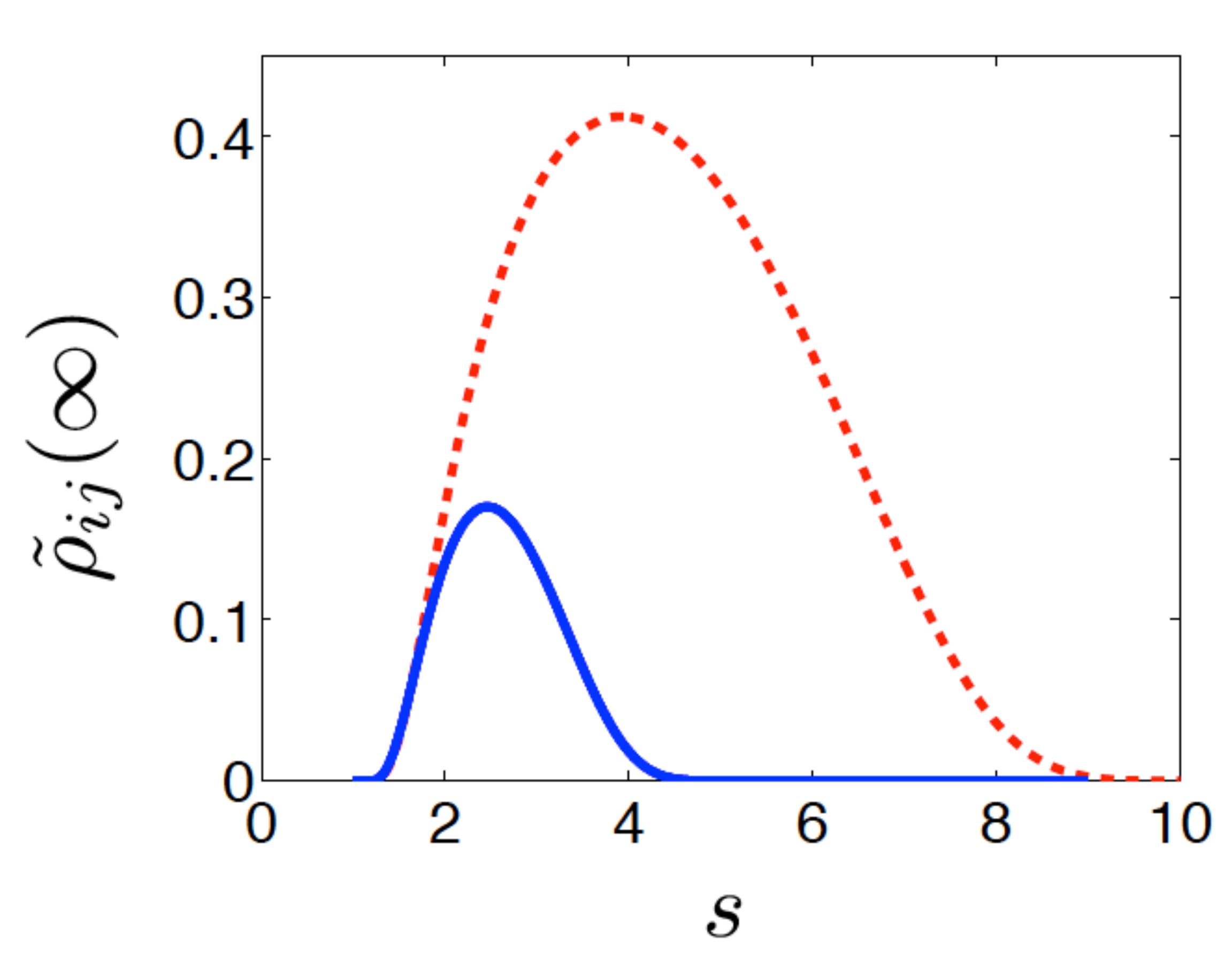}
\caption{(Color online) The stationary coherences $\tilde{\rho}_{ij}(\infty)$ as a function of the Ohmicity parameter $s$ for the softer (blue solid line) and harder (red dotted line) cutoff functions. Here we consider the system at zero temperature.   } 
\label{figure1}
\end{figure} 

{\it Information back-flow.} The dynamics of a qubit interacting with a purely dephasing environment can be characterized by looking at the non-monotonic behavior of certain quantities, and defining the amount of accessible information on the qubit when it is subjected to dephasing for a certain time interval $t$. A recently introduced quantifier of information back-flow (i.e., information flowing from the environment back to the system) is based on the time dependence of the quantum channel capacity \cite{Bogna}. This quantity gives a bound on the maximum rate at which quantum information can be reliably transmitted along a noisy quantum channel, and it is therefore of key importance in quantum communication protocols. In absence of reservoir memory effects (Markovian dynamics) the quantum channel capacity monotonically decreases in time due to the presence of dephasing noise. However, reservoir memory effects may lead to a non-monotonic behavior of the quantum channel capacity, or equivalently of the accessible information on the system. It is worth mentioning that, for the system here considered, the non-Markovianity measured in terms of partial increase of the quantum channel capacity coincides with previously introduced measures of non-Markovianity \cite{Rivas, Elsi}. More precisely, while the numerical values may change, the Markovian to non-Markovian crossover is the same and coincides with the divisibility or non-divisibility of the dynamical map. This is, in turn, reflected as monotonic or non-monotonic behavior of the decoherence factor $\Lambda(t)$. 

To formally quantify the memory effects associated with quantum channel capacity $Q(\phi_t)$, we calculate the following integral: 
\eq
\mathcal{N}_Q=\int_{\frac{dQ(\phi_t)}{dt}>0}\frac{dQ(\phi_t)}{dt}dt.
\label{B}
\eeq
With knowledge of the optimal qubit state and further simplifications the measure is analytical, encompassing the intervals $t\in(a_i,b_i)$  of information back-flow. 
\eq
\mathcal{N_Q}=\sum_i Q(b_i)-Q(a_i)
\eeq
where, 
\eq
Q(t)=1-H_2\left(\frac{1+e^{-\Lambda(t)}}{2}\right),
\eeq
with $H_2(.)$ the binary Shannon entropy. 
Moreover, it is straightforward to find the times $t\in(a_i,b_i)$ encapsulating non-monotonic intervals of $\Lambda(t)$ , i.e. times when $\frac{d\Lambda(t)}{dt}=0$. 

\begin{figure}[h]
\includegraphics[width=0.5\textwidth]{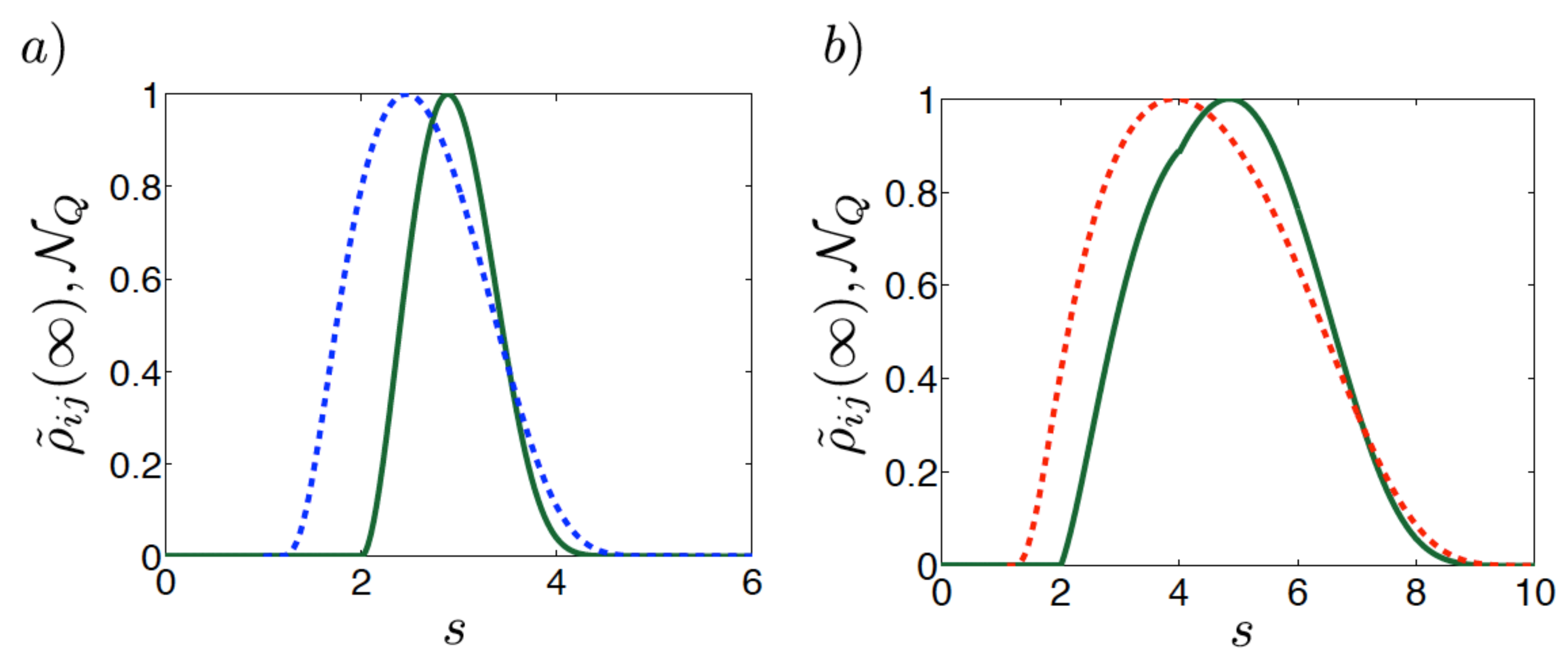}
\caption{(Color online) The non-Markovianity measure $\mathcal{N}_Q$ (green solid line) compared with the stationary coherences $\tilde{\rho}_{ij}(\infty)$ for a) the softer cutoff function (blue dotted line) and b) harder cutoff function (red dotted line) as a function of the Ohmicity parameter $s$. Note that both quantities are normalized to unity for easier comparison.} 
\label{figure3}
\end{figure} 

The necessary and sufficient conditions on the form of the spectrum to induce non-Markovian dynamics in the qubit, derived in Ref. \cite{tomipinja} for the exponential softer cutoff, are independent of the form of the cutoff function as long as it is monotonically decaying for $\omega \rightarrow \infty$. To understand this we recall that, for the Ohmic class, the necessary and sufficient condition for non-Markovianity coincides with the non-convexity of $g(\omega, T)$. It is easy to study the change of convexity in the two limiting cases of zero and high $T$. In the zero $T$- case, e.g., $g(\omega) \propto \omega ^{s-2} f(\omega, \omega_c)$. It is immediate to notice that a change of convexity in this case happens when passing from $s < 2$, for which $g(\omega)$ diverges in the origin, to $s> 2$, for which $g(\omega)$ vanishes in the origin,  independently on the specific form of the monotonically decaying $f(\omega, \omega_c)$. Exactly the same reasoning holds for the high $T$ case, with the critical value now being $s=3$. Finally, the generic $T$ case interpolates between the two. Hence we can conclude that the information back-flow depends on both $s$ and $T$ but not on the form of the cutoff function. Stated another way, it is the low frequency part of the spectrum that influences the presence or absence of information back-flow. 
 
{\it Conclusions. } We have examined the dependence of the stationary coherences and occurrence of information back-flow on the specific form of the spectral density function associated with the environment. The low frequency part of the spectrum rules both the occurrence of information back-flow and the occurrence of coherence trapping, independently on the form of the cutoff function. On the other hand, the high frequency part of the spectrum, i.e., form of the cutoff function (softer vs harder) rules the final value of the stationary coherences, when they exist. 

For the dephasing model considered in this article, we demonstrate two important features of non-Markovian dynamics. On one hand we show the existence of stationary coherences originating from the vanishing of $\omega g(\omega)$ in the limit $\omega\rightarrow 0$ when $s>1$, and on the other hand, information back-flow associated with the non-convexity of $g(\omega)$ when $s>2$. Stationary coherences are not dependent on information back-flow but are rather associated with the failure of the Markov approximation which predicts vanishing coherences for long times. As coherence trapping is not consistent with a simple Markovian dynamical description, we can classify it as a non-Markovian phenomenon. The maximum stationary coherence is achieved for $s>2$, i.e., in the non-Markovian region associated with a reversal of information from the environment to the system. Further, one can see from Fig. 2, that only when $s>2$ do the values of the stationary coherences for the softer and harder cutoff depart from each other as a result of entering into the non-Markovian regime.


\begin{thebibliography}{30}
\bibitem{lasercooled} D. Leibfriend \emph {et al}, Rev. Mod. Phys. {\bf75}, 281 (2003); H. H\"{a}ffner, C. F. Roos and R. Blatt, Phys. Rep. {\bf469}, 155 (2008).
\bibitem{opticallattices} I. Bloch \emph{et al}, Nature Phys. {\bf8}, 267 (2012). 
\bibitem{NVdiamond} M. W. Doherty \emph{ et al}, Phys. Rep, {\bf528}, 1 (2013).
\bibitem{dots} J. R. Petta \emph{et al} Science {\bf 309}, 2180-2184 (2005).
\bibitem{Susana} S. F. Huelga, \'{A}. Rivas and M. B. Plenio, Phys. Rev. Lett. {\bf 108}, 160402 (2012).
\bibitem{Nature} A. W. Chin \emph{et al} Nature Phys. {\bf 9}, 113 (2013). 
\bibitem{Susana2} A. W. Chin, S. F. Huegla and M. B. Plenio, Phys. Rev. Lett. {\bf 109}, 233601 (2012). 
\bibitem{Ruggero} R. Vasile et al, Phys. Rev. A {\bf 83}, 042321 (2001).
\bibitem{Teleportation} E.-M. Laine, H.-P. Breuer, J. Piilo,  arXiv:1210.8266 (2012). 
\bibitem{Bogna} B. Bylicka, D. Chru\'{s}ci\'{n}ski and S. Maniscalco, arXiv:1301.2585 (2013). 
\bibitem{luczka} J. Luczka, Physica A {\bf 167}, 919 (1990).
\bibitem{massimo} G. M. Palma, K.-A. Suominen, and A. K. Ekert, Proc. R. Soc. London, Ser. A {\bf 452}, 567 (1996).
\bibitem{reina} J. H. Reina, L. Quiroga and N. F. Johnson, Phys. Rev. A {\bf 65}, 032326 (2002).
\bibitem{Spectral} M. J. Biercuk, H. Uys, A.P. VanDevender, N. Shiga, W. M. Itano and J. J. Bollinger, Nature {\bf 458}, 996-1000 (2009).
\bibitem{Lorent}  T. E. Hodgson, L. Viola and I. D'Amico Phys. Rev. A {\bf 81}, 062321 (2010).
\bibitem{us} P. Haikka, S. McEndoo, G. De Chiara, G. M. Palma and S. Maniscalco, Phys. Rev. A {\bf 84}, 031602 (2011).
\bibitem{probes} P. Haikka, S. McEndoo, and S. Maniscalco, Phys. Rev. A {\bf 87}, 012127 (2013).
\bibitem{bang1} D. A. Lidar, arXiv:1208.5791 (2013); L. Viola, S. Lloyd and E. Knill, Phys. Rev. Lett. {\bf 83}, 4888 (1999).
\bibitem{bang2} E. Knill, R. Laflamme and L. Viola, Phys. Rev. Lett. {\bf 84}, 2525 (2000). 
\bibitem{J} U. Weiss, {\it Quantum Dissipative Systems} (World Scientific, Singapore, 1999).
\bibitem{tomipinja}P. Haikka, T. H. Johnson, and S. Maniscalco, Phys. Rev. A {\bf 87}, 010103(R) (2013).
\bibitem{note} We note that the $t \rightarrow \infty$ limit here should be understood as a long time limit but still such that $t < T_1$, i.e., heating is negligible.

\bibitem{Rivas} {\'A}. Rivas, S. F. Huelga and M. B. Plenio, Phys. Rev. Lett. {\bf105}, 050403 (2010).
\bibitem{Elsi} H.-P. Breuer, E.-M. Laine and J. Piilo, Phys. Rev. Lett. {\bf103}, 210401 (2009).

\bibitem{bp} H.-P.~Breuer and F.~Petruccione, The Theory of Open Quantum Systems, (Oxford Univ. Press, 2007).
\bibitem{L} A. Recati, P. O. Fedichev, W. Zwerger, J. von Delft, and P. Zoller, Phys. Rev. Lett. {\bf 94}, 040404 (2005).
\bibitem{B} M. Bruderer and D. Jaksch, New Jour. Phys. {\bf 8}, 87 (2006). 


\end{thebibliography}
\end{document}